\documentstyle[12pt]{article}

\newcommand{\al}{\alpha}
\newcommand{\ra}{\rightarrow}
\newcommand{\bt}{\beta}
\newcommand{\dlt}{\delta}
\newcommand{\Dlt}{\Delta}
\newcommand{\om}{\omega}
\newcommand{\gm}{\gamma}
\newcommand{\Gm}{\Gamma}
\newcommand{\prt}{\partial}
\newcommand{\lbd}{\lambda}

\begin{document}

\begin{center} 
{\Large{\bf Nuclear Spin Superradiance} \\ [5mm]

V.I. Yukalov} \\ [2mm]

{\it Bogolubov Laboratory of Theoretical Physics \\
Joint Institute for Nuclear Research, Dubna 141980, Russia}

\end{center}

\vskip 2cm

{\bf Tel:} 7 (096) 216 - 3947

\vskip 2mm

{\bf Fax:} 7 (096) 216 - 5084

\vskip 2mm

{\bf E-mail:} yukalov@thsun1.jinr.ru

\vskip 1cm

Vyacheslav I. Yukalov. b. 1946. M.Sc., 1970, Ph.D., 1974, Moscow, Dr.Hab.,
1999, Poznan, Dr.Sci., 2001, Moscow. First involved with NMR related 
research in 1984, after starting his work at JINR in Dubna, where together 
with an experimental group he took part in the first observation of pure 
nuclear spin superradiance. Then developed a microscopic theory of this
phenomenon. Received the First Prize of Joint Institute for Nuclear Research
for the discovery and theory of nuclear spin superradiance in 2001. Pubslished
250 papers and 4 books. Current research interests: developing the theory
of spin superradiance for including novel materials, and for finding new 
applications.

\newpage

\begin{center}
{\Large{\bf Nuclear Spin Superradiance} \\ [5mm]

Vyacheslav I. Yukalov} \\ [5mm]

{\it Bogolubov Laboratory of Theoretical Physics \\
Joint Institute for Nuclear Research, Dubna 141980, Russia}

\end{center}

\vskip 3cm

{\bf Table of Contents}

\begin{enumerate}

\item  
Introduction

\item 
Experimental Observation

\item
Basic Model

\item
Evolution Equations

\item
Nyquist Noise

\item
Incoherent Stage

\item
Transient Superradiance

\item
Pulsing Superradiance

\item
Hyperfine Interactions

\item
Radiation Intensity

\item
Applications

\end{enumerate}

\newpage

\section{Introduction}

Nuclear spins may exhibit several coherent phenomena, occurring at 
nuclear magnetic resonance frequencies, which have their direct 
counterpats in resonant atomic systems{\large$^1$}. For instance, 
nuclear free induction is an analog of atomic free induction and the
spin echo is an analog of the photon echo. These analogies are due to 
the fact that an ensemble of identical spins forms a collective of 
finite-level objects, similar to a resonant system of atoms or 
molecules. Such a finite-level nonequilibrium system can, under 
special conditions, behave as a collection of coherent radiators.

Similarly to the case of resonant atoms, one may distinguish two 
principally different ways of dealing with spin assemblies. One situation 
would be when spins are near their equilibrium state, with a 
nonequilibrium perturbation produced by a resonant alternating field. 
This is a typical situation of nuclear magnetic resonance{\large$^2$}. 
Another possibility could be if spins are initially prepared in a 
strongly nonequilibrium state, e.g. being polarized against a constant 
external magnetic field. In the latter case, the polarized spins resemble
a system of inverted resonant atoms.

One of the most interesting effects exhibited by a system of inverted 
atoms is superradiance, when the radiation intensity is approximately 
proportional to the number of atoms squared, while the duration of a 
superradiant pulse is inversely proportional to this number. The 
possibility of {\it atomic superradiance} was predicted by 
Dicke{\large$^3$}, and nowadays it is well studied both theoretically 
and experimentally. There exists a vast literature on the subject, whose 
description can be found in the recent books{\large$^{4,5}$}. A natural 
question that arises is: If atomic and spin systems are so similar, 
then could a kind of {\it spin superradiance} be realized?

For spins to behave coherently, there must exist a cause correlating 
their motion. In the case of atoms, such a correlating mechanism is 
caused by the photon exchange through the common radiation field. This 
exchange results in the formation of effective interatomic correlations 
collectivizing atomic radiation. For spins, however, their 
magneto-dipole radiation is too weak to develop noticeable spin 
correlations. This concerns both nuclear as well as electron spins. Such 
photon-exchange interactions are negligible as compared to disordering 
dipolar spin interactions. How then might the spin motion be collectivized?

It is, of course, possible to force an ensemble of spins to develop 
coherence by imposing an initial condition of longitudinal magnetization, 
and using an rf pulse to produce a macroscopic transverse magnetization.
But this would result in the standard free nuclear induction, with the 
coherence being lost during the dephasing time $T_2$. However, free 
nuclear induction, although being a coherent process, is not 
superradiance with one of the main characteristic features of a short 
radiation time of a superradiant pulse $\tau_p\ll T_2$,
shorter than the transverse dephasing time. To make the superradiant 
pulse time $\tau_p$ so short, some internal nonlinear correlating 
mechanism has to be involved.

Mutual spin correlations can arise owing to a feedback field formed by 
a resonant electric circuit coupled to the spin system and tuned to the 
Zeeman transition frequency of spins. The role of such a coupling in 
magnetic resonance experiments has been analysed by Bloembergen and 
Pound{\large$^6$}. They showed that, in the presence of an electric 
resonant circuit, coupled to a spin system, the signal of nuclear 
induction can be damped in a time much smaller than $T_2$. Since this 
shortening of the induction damping time is due to collective effects, 
caused by the spin coupling through the resonator feedback field, the
resulting process may be termed {\it collective induction}. This effect
has been observed in a number of substances, as has been reviewed{\large$^7$}.
The process already possesses one of the prerequisits of superradiance, 
i.e. a short radiation time $\tau_p< T_2$. Usually, the induction signal 
starts at $t=0$, having there its maximum intensity, while a superradiant 
pulse is always separated from the time origin by a delay time $t_0>0$. 
The signal of collective induction can also be peaked at a delay time well 
separated from $t=0$. However, there is a principal difference between 
collective induction and superradiance: In collective induction, coherence 
is induced by external sources, while in superradiance coherence 
develops in a self-organized way owing to internal correlations. The 
{\it self-organized, spontaneous}, nature of superradiance is one of its 
most important features{\large$^{3-5}$}. In general, one includes into the 
class of superradiant the processes that are triggered by initial external 
pulses, but with a compulsory requirement that the imposed initial coherence 
be very weak, playing just the role of a trigger. Thus, collective induction, 
though being a collective coherent process, is not yet superradiance.

The principal characteristics of superradiance, which apply both to atomic 
as well as to spin systems, can be summarized as follows:

\begin{enumerate}

\item Superradiance is collective, coherent radiation by an ensemble of 
radiators.

\item It is a spontaneous process developing in a self-organized way.

\item The maximal intensity of a superradiant burst is proportional to the 
number of radiators in the power larger than one.

\item The duration of a superradiant pulse is essentially shorter than the 
dephasing tine $T_2$.

\item A superradiant pulse has a peak at a finite delay time.

\end{enumerate}

In atomic systems, one may distinguish different types of superradiance, 
all having the same characteristic features, including a short pulse time 
$\tau_p< T_2$ and finite delay time $t_0>0$. First of all, superradiance 
can occur as a transient process or as a lasting repeated effect, depending 
on whether there is no external nonresonant pumping permanently applied to 
the system or, respectively, there exists such a pumping supporting atomic 
inversion. In the first case, the system is prepared in an inverted state, 
after which no nonresonant pumping is involved. Then, if the relaxation 
process is gently promoted at $t=0$ by an external pulse imposing weak 
initial coherence on the system, {\it triggered superradiance} may develop. 
When no initial coherence is thrust on the system, but superradiance appears 
as a completely self-organized effect starting from a purely incoherent 
state, then this is named {\it pure superradiance}. In the case, when the 
system is subject to the action of a nonresonant pumping constantly supporting 
atomic inversion, the regime of {\it pulsing superradiance} may arise, with 
a long series of superradiant bursts.

It has been of great interest to find out if a kind of spin 
superradiance, occurring in spin systems similarly to this phenomenon
in atomic assemblies, could be realized experimentally and, if so, how
should it be described theoretically. Despite several similarities 
between spin and atomic ensembles, they are, nevertheless, rather 
different. For example, spontaneous radiation, starting the process 
of pure atomic superradiance, is absent in spin assemblies. Therefore, 
one of the most intriguing questions has been if pure spin superradiance
can exist and, if so, what would be its origin.

The phenomenon of spin superradiance, being realized, could be of 
interest by its own and, in addition, it could be employed for several 
applications, as is discussed in section 11. For example, these 
applications could include:

\vskip 2mm

(i) {\it Investigation of materials characteristics} by measuring the 
relaxation parameters that are specific for spin superradiance and do not 
arise in other types of spin relaxation.

\vskip 2mm

(ii) {\it Fast repolarization of targets} used in various scattering 
experiments of high-energy physics.

\vskip 2mm

(iii) {\it Construction of spin masers} producing coherent radiation at 
radiofrequencies.

\vskip 2mm

(iv) {\it Creation of sensitive detectors} of weak external pulses, 
based on the mechanism of triggered spin superradiance.

\vskip 2mm

(v) {\it Method of information processing}, derived from the feasibility 
of regulating the number of and the intervals between superradiant bursts
in the regime of punctuated spin superradiance.

\section{Experimental Observation}

Historically, the regime of {\it pulsing spin superradiance} was observed 
first{\large$^{8-10}$}. This was done for a ruby crystal Al$_2$O$_3$, with 
the Cr$^{3+}$ paramagnetic admixture. The active nuclei were $^{27}$Al, with 
spins $I=5/2$. If the ruby crystal is oriented in an external magnetic field 
such that a fully resolved quadrupole structure of its five $\Delta m=\pm 1$ 
NMR transitions can be observed, and a resonant circuit is tuned to a 
selected NMR line, then $^{27}$Al spins form a fictitious two-level system. 
In experiments, the resonant circuit was tuned to the central 
$\left\{-\frac{1}{2},\frac{1}{2}\right\}$ line. The density of $^{27}$Al 
nuclei was $\rho_{Al}=4.4\times 10^{22}$ cm$^{-3}$ and that of Cr$^{3+}$, 
$\rho_{Cr}=8.6\times 10^{18}$ cm$^{-3}$. The measurements were performed 
in a static magnetic field of about $B_0=1.1$ T, which for the $^{27}$Al 
magnetogyric ratio $7\times 10^7$ s$^{-1}$T$^{-1}$ makes the NMR frequency 
$8\times 10^7$ Hz. The temperature range was from $1.6$ to 
$4.2$ K. The ringing time of the resonant circuit was $\tau=10^{-6}$ s, 
and the quality factor $Q$ was varied between $60$ and $200$. The coil 
filling factor was $\eta=0.55$. The transverse spin damping time was 
$T_2=3\times 10^{-5}$ s. The spin population inversion of the $^{27}$Al 
nuclear spins was achieved by dynamic nuclear polarization, by means of 
powerful microwave radiation supplied to the sample in the vicinity of 
a selected Cr$^{3+}$ electron spin resonance line. The corresponding 
longitudinal spin pumping time was $T_1^*=0.1$ s. The emission of 
a long train of superradiant bursts was observed, with the delay 
time of the first pulse being $t_0=2\times 10^{-3}$ s and its duration 
being $\tau_p=2.6\times 10^{-5}$ s in the case of the quality factor 
$Q=60$, while $\tau_p=8\times 10^{-6}$ s for $Q=200$.

The first observation of {\it pure spin superradiance} was accomplished
in Dubna{\large$^{11,12}$}. A dielectric propanediol C$_3$H$_8$O$_2$,
with a paramagnetic admixture of Cr$^{5+}$, was used as an active 
substance. This material possesses a high concentration of 
protons, with density $\rho_H=4\times 10^{22}$ cm$^{-3}$. The proton
spins $I=1/2$ played the role of radiators. The admixture of Cr$^{5+}$,
with the density $\rho_{Cr}=1.8\times 10^{20}$ cm$^{-3}$, was employed
for the purpose of dynamic nuclear polarization of proton spins. 
Experiments with the same material propanediol were repeated in Saint 
Petersburg{\large$^{13}$}. In all these experiments{\large$^{11-13}$},
electric circuits were used as resonators, with quality 
factors $Q$ between 100 and 600. The filling factor was $\eta=0.6$. The
volume of samples varied from 0.5 cm$^3$ to 12 cm$^3$. For external 
static magnetic fields $B_0$ in the interval between $0.5$ T and $2.64$ T, 
with the proton magnetogyric ratio $2.675\times 10^8$ s$^{-1}$T$^{-1}$, 
the NMR proton frequencies were between $1.3\times 10^8$ Hz and 
$7.1\times 10^8$ Hz. The experiments were carried out at low temperatures 
of 0.05 K to 0.1 K. The inverted proton polarization reached 90$\%$. Cooling 
of the sample resulted in strong suppression of the nuclear spin-lattice 
relaxation. The related longitudinal relaxation time was $T_1=1.8\times 10^5$ 
s at $B_0=0.5$ T and $T_1=1.8\times 10^6$ s at $B_0=2.64$ T. The transverse 
dephasing time was $T_2=0.85\times 10^{-5}$ s. After preparing an inverted 
sample, with the proton polarization directed against an external static 
magnetic field $B_0$, the latter was scanned with the velocity about 
$5\times 10^{-3}$ Ts$^{-1}$. When in the process of this slow adiabatic 
scanning the field value $B_0$ reached that one for which the proton NMR 
frequency coincided with the circuit natural frequency, the resonance 
condition was met. Then a powerful superradiant burst was produced, with 
the duration time about $\tau_p=1.3\times 10^{-6}$ s.

Similar experiments, observing pure spin superradiance, were accomplished
in Bonn{\large$^{14}$}, with the target materials butanol C$_4$H$_9$OH 
and ammonia NH$_3$. These materials are also rich with protons, with density 
$\rho_H\approx 3\times 10^{23}$ cm$^{-3}$. By means of dynamic nuclear 
polarization, proton polarizations of up to 99$\%$ were achieved. The samples 
were cooled to low temperatures, the spin-lattice relaxation time was 
$T_1\approx 3.6\times 10^4$ s for protons in butanol and $T_1\approx 10^5$ s 
for protons in ammonia. The resonant electric circuit, with the quality factor 
$Q=33$, had a resonance frequency of $1.6\times 10^8$ Hz.

In different experiments, slightly different setups were employed. A typical 
arrangement{\large$^{11,12}$} is shown in Fig. 1. The characteristic 
superradiant pulse{\large$^{11,12}$} is presented in Fig. 2.

A series of numerical simulations were accomplished{\large$^{15,16}$},
which could be considered as computer experiments modelling the 
behaviour of spins in conditions typical of experimental observations. 
These computer simulations confirmed the existence of both pure as well 
as triggered spin superradiance, being in good agreement with experiments.

\section{Basic Model}

Since the phenomenon of spin superradiance exists in nature, it is necessary 
to develop its theoretical description. The simplest idea would be to invoke 
for this purpose the Bloch equations. However, these are designed so that 
they treat the magnetization of a sample as a macroscopic vector, which 
presupposes the existing coherence from the very beginning. If at the initial 
time coherence of spins is absent, the Bloch equations would never display it.
Therefore, these equations can describe only triggered spin superradiance, 
when coherence is thrust upon spins at the initial time by assuming the 
presence of nonzero transverse magnetization. But pure spin superradiance, 
being a self-organized coherent process cannot in principle be treated by 
the Bloch equations. A discussion of this problem as well as the related 
references can be found in a review{\large$^{17}$}. In order to be able to 
describe all possible regimes of spin dynamics, it is necessary to resort to 
microscopic models. A theory of {\it nuclear spin superradiance}, based on 
realistic Hamiltonians typical of nuclear spin assemblies in condensed 
matter{\large$^{1,2,18}$} has been previously developed{\large$^{19-24}$}.

A system of $N$ nuclear spins, enumerated by an index $i=1,2,\ldots,N$, 
is characterized by their spin operators $\hat{\bf I}_i$. The Hamiltonian
of nuclear spins in a solid can be written as a sum
\begin{equation}
\label{1}
\hat H_{nuc} =\sum_i \hat H_i +\frac{1}{2}\sum_{i\neq j} \hat H_{ij}
\end{equation} 
of the Zeeman term and of a part due to many-body spin interactions. The 
Zeeman term contains
\begin{equation}
\label{2} 
\hat H_i = - \hbar \gm_n{\bf B}\cdot\hat{\bf I}_i \; ,
\end{equation}
where ${\bf B}$ is the total magnetic field and $\gm_n$ is the nuclear 
magnetogyric ratio. The total magnetic field
\begin{equation}
\label{3} 
{\bf B} = B_0{\bf e}_z + (B_1 + H){\bf e}_x
\end{equation}
consists of a static magnetic field $B_0$ along the $z$-axis and of 
a transverse field being a sum of an external field $B_1$ and of a 
field $H$ produced by an electric coil which the sample is immersed 
into. The corresponding orientation of the coordinate axes with 
respect to the sample is explained in Fig. 3. The static magnetic field
is directed so that 
\begin{equation}
\label{4} 
\gm_n B_0 < 0 \; .
\end{equation}
In general, the nuclear magnetogyric ratio can be positive as well as 
negative. If it is positive, then inequality (4) tells that the static
field is negative. In this case, the initial polarization of the sample
nuclei is called inverted if it is positive, that is directed 
against the static magnetic field.

The basic force acting between nuclear spins is due to the dipolar interactions
which may be described by the Hamiltonian
\begin{equation}
\label{5} 
\hat H_{ij} =\sum_{\al\bt} D_{ij}^{\al\bt} \hat I_i^\al \hat I_j^\bt \; ,
\end{equation}
where the upper indices $\al$ and $\bt$ denote the coordinate 
components $\al,\bt=x,y,z$, while
\begin{equation}
\label{6} 
D_{ij}^{\al\bt} = \frac{\hbar^2\gm_n^2}{r_{ij}^3} \left ( \dlt_{\al\bt} -
3n_{ij}^\al n_{ij}^\bt\right )
\end{equation}
is the dipolar tensor, in which
$$
r_{ij} \equiv |{\bf r}_{ij}| \; , \qquad n_{ij} \equiv 
\frac{{\bf r}_{ij}}{r_{ij}}\; , \qquad {\bf r}_{ij} \equiv {\bf r}_i -
{\bf r}_j \; .
$$
The dipolar tensor satisfies the equalities 
\begin{equation}
\label{7} 
\sum_\al D_{ij}^{\al\al} = 0 \; , \qquad \sum_{j(\neq i)} D_{ij}^{\al\bt}
= 0 \; ,
\end{equation}
of which the first is exact and the second is asymptotically valid 
for a macroscopic sample.

For practical purpose, it is convenient to work with the ladder spin operators 
$\hat I_j^\pm \equiv \hat I_j^x\pm \hat I_j^y$, called the raising and lowering 
spin operators, respectively. Then, with the help of the notation
$$ 
a_{ij} \equiv D_{ij}^{zz} \; , \qquad
b_{ij} \equiv \frac{1}{4}\left ( D_{ij}^{xx} - 2i D_{ij}^{xy} -
D_{ij}^{yy}\right ) \; ,
$$
\begin{equation}
\label{8} 
c_{ij} \equiv \frac{1}{2}\left ( D_{ij}^{xz} - i D_{ij}^{yz}\right )\; ,
\end{equation}
the Zeeman term (2), for the total field (3), becomes 
\begin{equation}
\label{9} 
\hat H_i = -\hbar\gm_n B_0 \hat I_i^z -\; \frac{1}{2}\; \hbar \gm_n\left ( 
B_1 + H \right ) \left ( \hat I_i^+ + \hat I_i^- \right )
\end{equation}
and the dipolar interaction (5) takes the form
$$
\hat H_{ij} = a_{ij}\left ( \hat I_i^z \hat I_j^z -\; \frac{1}{2}\hat I_i^+
\hat I_j^- \right ) + b_{ij}\hat I_i^+\hat I_j^+ + b_{ij}^*\hat I_i^- 
\hat I_j^- +
$$
\begin{equation}
\label{10} 
+ 2c_{ij} \hat I_i^+\hat I_j^z + 2c_{ij}^*\hat I_i^-\hat I_j^z \; .
\end{equation}
Equations (9) and (10) will be used below to derive the evolution equations.

The electric circuit, coupled with the spin sample, is characterized 
by resistance $R$, inductance $L$, and capacity $C$. The coil, 
surrounding the sample, has $n$ turns of cross-section area $A_c$ over
a length $l$. The electric current $j$ of the circuit is given by 
the Kirchhoff equation
\begin{equation}
\label{11} 
L\; \frac{dj}{dt} + Rj + \frac{1}{C} \; \int_0^t j(t')\; dt' =
\tilde E - \; \frac{d\Phi}{dt} \; ,
\end{equation}
in which $\tilde E$ is an electromotive force, if any, and $\Phi$ is 
a magnetic flux
\begin{equation}
\label{12} 
\Phi=\frac{4\pi}{c}\; n A_c \eta M_x
\end{equation}
formed by the $x$-component of the magnetization density
\begin{equation} 
\label{13}
M_x =\frac{\hbar\gm_n}{V} \sum_i <\hat I_i^x> \; .
\end{equation}
Here the brackets $<\ldots>$ imply statistical averaging. The filling factor 
$\eta$ is approximately equal to $\eta\approx V/V_c$, where $V$ is the sample 
volume, while $V_c\equiv A_c l$ is the coil volume.

The electric current in the circuit is formed by moving transverse 
spins. In its turn, the current, circulating over the coil, produces 
a feedback magnetic field
\begin{equation} 
\label{14}
H = 4\pi \frac{nj}{cl} \; .
\end{equation}
Hence, the Kirchhoff equation (11) can be rewritten for the feedback field 
(14). To this end, it is useful to employ the notation for the natural circuit 
frequency
\begin{equation} 
\label{15}
\om \equiv \frac{1}{\sqrt{LC}}  \qquad \left ( L \equiv 4\pi\;
\frac{n^2A_c}{c^2l}\right )
\end{equation}
and for the circuit ringing time
\begin{equation} 
\label{16}
\tau \equiv \frac{1}{\gm} \qquad \left ( \gm \equiv \frac{R}{2L}\right ) \; .
\end{equation}
The related circuit damping
\begin{equation} 
\label{17}
\gm =\frac{\om}{2Q} \qquad \left ( Q \equiv \frac{\om L}{R}\right )
\end{equation}
is connected with the quality factor $Q$. Employing these notations, 
together with that for the reduced electromotive force
\begin{equation} 
\label{18}
h \equiv \frac{c\tilde E}{nA_c\gm} \; ,
\end{equation}
one obtains from the Kirchhoff equation (11) the result
\begin{equation} 
\label{19}
\frac{dH}{dt} + 2\gm H  + \om^2 \int_0^t H(t')\; dt' =
\gm h -4\pi\eta \; \frac{dM_x}{dt}
\end{equation}
for the feedback magnetic field (14). 

The feedback equation (19) can be presented in a form that extremely useful 
to exploit for analysing the evolution equations{\large$^{20}$}. By envolving 
the method of Laplace transforms and introducing the transfer function
$$
G(t) =\left ( \cos\tilde \om t -\; \frac{\gm}{\tilde\om}\;
\sin\tilde\om t\right ) e^{-\gm t} \; ,
$$
with $\tilde\om\equiv\sqrt{\om^2-\gm^2}$, one may present eq. (19) as 
the integral
\begin{equation} 
\label{20}
H =\int_0^t G(t-t') \left [ \gm h(t') - 4\pi\eta\dot{M}_x(t')
\right ]\; dt' \; ,
\end{equation}
where the dot over $\dot{M}_x$ means, as usual, time derivative.

\section{Evolution Equations}

The equations of motion for the nuclear spin operators are given by the 
corresponding Heisenberg equations. From these one aims at deriving the 
equations for the following statistical averages. The function
\begin{equation}
\label{21}
u \equiv \frac{1}{I}\; <\hat I_i^->
\end{equation}
defines the rotation of transverse spin components. The degree of coherence 
in this rotation is described by
\begin{equation}
\label{22}
w\equiv \frac{1}{I^2}\; < \hat I_i^+><\hat I_i^->\; = |u|^2 \; .
\end{equation}
And the average
\begin{equation}
\label{23}
s\equiv \frac{1}{I}\; <\hat I_i^z> 
\end{equation}
gives the longitudinal spin polarization, which is analogous to the population
difference in optic systems. Keeping in mind that wavelengths at 
radio-frequencies are much larger than mean distances between spins and 
usually even essentially larger than linear sizes of a sample, one may 
employ the uniform approximation, assuming that the functions (21) to (23) 
do not depend on spatial variables. 

In order to obtain the evolution equations for the quantities (21) to (23), 
one averages the Heisenberg equations for the corresponding spin operators. 
In doing so, one encounters the known problem of the resulting equations being
not closed but containing, in addition to $u$, $w$, and $s$, also binary 
averages of spin operators. The standard way of closing these equations would 
be by employing the mean-field decoupling, when the binary average 
$<\hat I_i^\al\hat I_j^\bt>$, with $i\neq j$, is factorized onto the product 
$<\hat I_i^\al><\hat I_j^\bt>$. Such a decoupling, however, ends with the 
Bloch-type equations having the same deficiency of never exhibiting 
coherence if at the initial time $w(0)=0$. This is because the mean-field 
decoupling completely neglects spin correlations leading to local spin 
fluctuations. Taking account of such fluctuations is a necessary prerequisit 
for describing pure spin superradiance.

Another possibility of closing the hierarchy of spin evolution equations 
could be by writing additional equations for the binary correlators 
$<\hat I_i^\al\hat I_j^\bt>$ and by decoupling only the higher-order 
correlators, retaining untouched the binary ones. Unfortunately, this 
drastically increases the number of equations and so complicates the 
problem that it becomes practically untreatable even for equilibrium 
magnets{\large$^{25}$}. The more so for nonequilibrium systems presented 
by nonlinear differential equations.

To render the set of spin evolution equations closed so that to keep the 
problem treatable and, at the same time, to take into account local spin 
fluctuations, one can apply the {\it method of stochastic 
decoupling}{\large$^{19-21,24}$}. The main idea of the latter is to separate 
in the evolution equations the terms describing long-range correlations from 
the terms related to short-range fluctuations. Employing the method of 
restricted traces, one may define two types of statistical averages, one 
incorporating only long-range correlations and another one involving 
short-range fields that are treated as stochastic variables. In what follows, 
the first type of averaging is denoted by the single angle brackets $<\dots>$, 
while the second type, by the double brackets $\ll\ldots\gg$. Then, the binary 
spin correlators are presented as
\begin{equation}
\label{24}
<\hat I_i^\al\;\hat I_j^\bt>\; = \; <\hat I_i^\al><\hat I_j^\bt> \qquad
(i\neq j) \; ,
\end{equation}
where the averaging does not touch stochastic variables which enter the
evolution equations in the following linear combinations:
$$
\xi_0 \equiv \frac{1}{\hbar}\sum_{j(\neq i)} \left ( a_{ij}\hat I_j^z + 
c_{ij}\hat I_j^+ + c_{ij}^*\hat I_j^-\right ) \; ,
$$
\begin{equation}
\label{25}
\xi \equiv -\; \frac{i}{\hbar} \sum_{j(\neq i)} \left ( \frac{1}{2}\;
a_{ij}\hat I_j^- - 2b_{ij}\hat I_j^+ - 2 c_{ij}\hat I_j^z \right ) \; ,
\end{equation}
with the coefficients defined in eq. (8). Stochastic variables (25) describe 
local fluctuating fields which act on neighbour spins forcing them to move. 
The existence of such fluctuating fields should lead to the appearance of 
dipolar dynamic broadening{\large$^{26}$}. To completely define the problem, 
it is necessary to prescribe the rules of calculating stochastic averages over 
the random variables (25). This can be done by considering the stochastic 
variables (25) as describing a white-noise process characterized by the 
stochastic averages
$$
\ll\xi_0(t)\gg \; = \ll \xi(t)\gg \; = 0 \; , \qquad
\ll \xi_0(t)\xi_0(t')\gg \; = 2\Gamma_3\dlt (t-t') \; ,
$$
\begin{equation}
\label{26}
\ll\xi_0(t)\xi(t')\gg \; = \; \ll \xi(t)\xi(t')\gg \; = 0 \; , \quad
\ll \xi^*(t)\xi(t')\gg \; = 2\Gamma_3\dlt(t-t') \; ,
\end{equation}
in which $\Gamma_3$ is the width of {\it dynamic broadening} due to local 
dipole interactions of nuclei. This is a sort of inhomogeneous broadening 
existing additionally to the homogeneous broadening yielding the longitudinal,
$\Gamma_1\equiv T_1^{-1}$, and transverse, $\Gamma_2\equiv T_2^{-1}$, 
relaxation widths.

To obtain the resulting evolution equations in a compact form, it is convenient 
to introduce the notation
\begin{equation}
\label{27}
f \equiv -i\gm_n \left (B_1 + H\right ) + \xi
\end{equation}
for an effective force acting on a spin. The transverse external magnetic 
field $B_1$ may, in general, contain a static as well as an alternating term, 
as
\begin{equation}
\label{28}
B_1 = h_1 + h_2\cos\om t \; ,
\end{equation}
where, for simplicity, the frequency of the alternating field is assumed to 
be in resonance with the frequency of the electric circuit. Recall that the 
magnetic field $H$, produced by the coil, is caused by a feedback field and, 
possibly, by a field of an electromotive force, which may be presented as
\begin{equation}
\label{29}
h(t) = h_0\cos\om t \; .
\end{equation}
Finally, the NMR frequency is denoted by
\begin{equation}
\label{30}
\om_0 \equiv \;  | \gm_n B_0| \; .
\end{equation}
Then the evolution equations for the functions (21) to (23) acquire the form
\begin{equation}
\label{31}
\frac{du}{dt} = - i (\om_0 + \xi_0 - i\Gamma_2) u + fs \; ,
\end{equation}
\begin{equation}
\label{32}
\frac{dw}{dt} = - 2\Gamma_2 w + ( u^*f + f^* u ) s \; ,
\end{equation}
\begin{equation}
\label{33}
\frac{ds}{dt} = - \; \frac{1}{2} ( u^*f + f^* u ) - \Gamma_1( s-\zeta) \; ,
\end{equation}
where $\zeta$ is an average stationary value for a $z$-component of a spin. 
Generally, $-1\leq\zeta\leq 1$. When there is no external pumping, then 
$\zeta=-1$. In the presence of pumping, e.g. by means of dynamic nuclear 
polarization, the pumping parameter $\zeta>-1$ and can reach the value 
$\zeta=1$. Equations (31) to (33) compose a nonlinear system of stochastic 
differential equations describing all dynamic properties of nuclear spins.

The system of evolution equations (31) to (33) looks yet rather complicated. 
Fortunately, there are several small parameters that allow one to simplify 
the consideration by employing the {\it scale separation 
approach}{\large$^{19-21,27}$} which is a generalization of the 
Krylov-Bogolubov averaging technique{\large$^{28}$} to stochastic and 
partial differential equations. The existing small parameters are connected 
with the relatively small values of the relaxation widths $\Gamma_1$ and 
$\Gamma_2$ as compared to the NMR frequency (30), so that
\begin{equation}
\label{34}
\frac{\Gamma_1}{\om_0} \; \ll 1 \; , \qquad 
\frac{\Gamma_2}{\om_0} \; \ll 1 \; .
\end{equation}
Clearly, the quantity
\begin{equation}
\label{35}
\Gamma_0 \equiv \pi\eta\rho_n\gm_n^2 \hbar I \qquad
\left ( \rho_n \equiv \frac{N}{V}\right )
\end{equation}
is also small with respect to $\om_0$, as well as the dynamic broadening 
width $\Gamma_3$, that is
\begin{equation}
\label{36}
\frac{\Gamma_0}{\om_0} \; \ll 1 \; , \qquad 
\frac{\Gamma_3}{\om_0} \; \ll 1 \; .
\end{equation}
All amplitudes of the applied transverse fields are assumed to the small too,
which implies that the values
\begin{equation}
\label{37}
\nu_0 \equiv \gm_n h_0 \; , \qquad \nu_1 \equiv \gm_n h_1 \; , \qquad
\nu_2 \equiv \gm_n h_2 
\end{equation}
satisfy the inequalities
\begin{equation}
\label{38}
\frac{|\nu_0|}{\om_0} \; \ll 1 \; , \qquad 
\frac{|\nu_1|}{\om_0} \; \ll 1 \; , \qquad  \frac{|\nu_2|}{\om_0} \; \ll 1 \; .
\end{equation}
The electric circuit, coupled to the spin system, is supposed to be of good 
quality, having a high quality factor, which means that the ringing width is 
small compared to the natural circuit frequency,
\begin{equation}
\label{39}
\frac{\gm}{\om} \; \ll 1 \qquad ( Q\gg 1) \; .
\end{equation}
The last, though not the least, is that the electric circuit has to be tuned 
to the NMR frequency (30), hence the resonance condition for the detuning 
$\Dlt$ must be valid:
\begin{equation}
\label{40}
\frac{|\Dlt|}{\om_0} \; \ll 1 \qquad (\Dlt\equiv \om -\om_0 ) \; .
\end{equation}
In this way, the circuit plays the role of a resonator.

The existence of the small parameters makes it possible, by analysing the 
right-hand sides of the evolution equations (31) to (33), to conclude that 
the function $u$ has to be classified as fast, as compared to the slow 
functions $w$ and $s$. Conversely, the functions $w$ and $s$ are 
quasi-invariants with respect to $u$. In addition, one may derive an 
explicit expression for the resonator magnetic field $H$ by iterating 
its integral representation with the solution of eq. (31), where only the 
main term in the right-hand side is retained. This iteration gives
\begin{equation}
\label{41}
\gm_n H = i\al (u - u^*) + \bt\cos\om t \; ,
\end{equation}
where the first term, with the coupling function
\begin{equation}
\label{42}
\al \equiv \frac{\Gamma_0\om_0}{\gm}\left ( 1 - e^{-\gm t} \right ) \; ,
\end{equation}
is caused by the feedback coupling of spins with the resonator, and the second
term, with the coupling
\begin{equation}
\label{43}
\bt \equiv \frac{\nu_0}{2}\left ( 1 - e^{-\gm t} \right ) \; ,
\end{equation}
is due to an electromotive force. The time dependence of the coupling 
functions (42) and (43) describes the retardation in the resonator action 
on spins. Expressions (42) and (43) are written here for the case of small 
detuning, such that $|\Dlt|\ll\gm$. The latter assumption is not principal 
but is employed solely for the simplification of formulas. Relation (41) holds
true for any detuning satisfying the resonance condition (40).

Substituting relation (41) in eqs. (31) to (33) and using the notation
\begin{equation}
\label{44}
f_1 \equiv -i\nu_1 - i(\nu_2 +\bt)\cos\om t + \xi \; ,
\end{equation}
one comes to the evolution equations
\begin{equation}
\label{45}
\frac{du}{dt} = - \left [ i(\om_0 +\xi_0) + \Gamma_2 -\al s\right ] u +
f_1 s -\al s u^* \; ,
\end{equation}
\begin{equation}
\label{46}
\frac{dw}{dt} = - 2(\Gamma_2 -\al s)w + ( u^* f_1 + f_1^* u) s -
\al s \left [ (u^*)^2 + u^2\right ]  \; ,
\end{equation}
\begin{equation}
\label{47}
\frac{ds}{dt} = -\al w - \; \frac{1}{2}\left ( u^* f_1 + f_1^* u \right ) -
\Gamma_1 ( s-\zeta)  + \frac{1}{2}\; \al \left [ (u^*)^2 + u^2\right ]  \; .
\end{equation}
Following further the scale separation approach, eq. (45) for the fast 
function $u$ can be solved, with $w$ and $s$ being quasi-invariants. The 
solution obtained is substituted in the right-hand sides of eqs. (46) and 
(47), which are averaged over the period $2\pi/\om_0$ of fast oscilaltions 
as well as over stochastic variables defined in eq. (25). Then, introducing 
the effective attenuation width
\begin{equation}
\label{48}
\Gamma_{eff} \equiv \Gamma_3 -\al\; \frac{\nu_1^2}{\om_0^2}\; s - \; 
\frac{\nu_1(\nu_2 +\bt)\Gamma}{2\om_0^2}\; e^{-\Gamma t} +
\frac{(\nu_2+\bt)^2}{4\Gamma}\left ( 1 - e^{-\Gamma t}\right ) \; , 
\end{equation}
where
$$
\Gamma \equiv \Gamma_2 - \al s + \Gamma_3 \; ,
$$
reduces eqs. (46) and (47) for slow functions to the evolution equations
\begin{equation}
\label{49}
\frac{dw}{dt} = - 2 (\Gamma_2 - \al s ) w + 2 \Gamma_{eff}s^2 \; ,
\end{equation}
\begin{equation}
\label{50}
\frac{ds}{dt} = - \al w - \Gamma_{eff}s - \Gamma_1 ( s -\zeta ) \; .
\end{equation}
In the form of the widths (48), it is again assumed that the detuning is 
small, such that $|\Dlt|\ll|\gamma|\ll\om_0$, which is not principal but 
just simplifies combersome expressions.

The evolution equations (49) and (50) are the main equations describing the 
dynamics of strongly nonequilibrium nuclear spins. The analysis of these 
equations makes it possible to study various regimes of spin motion.

\section{Nyquist Noise}

The problem of principal importance is: What is the origin of pure spin 
superradiance? In other words, what initiates the motion of spins when no 
coherence is thrust upon the system at $t=0$ and there are no external fields 
pushing spins in the transverse direction?

Keeping in mind the analogy with atomic assemblies, one may remember that in 
an inverted system of atoms the relaxation process begins with atomic 
spontaneous radiation, which is a quantum process. After the seed radiation 
field appears in the system, atomic correlations start arizing through the 
interatomic photon exchange. As soon as these correlations become sufficiently
intensive, coherence develops. Then, the quantum stage of spontaneous emission 
changes for the coherent stage, when atoms are correlated and emit coherently, 
which results in superradiance.

The collectivization of spins can be produced by means of the resonator 
feedback field. But for this field to arise, the spins, first, have to start 
their motion. Spontaneous emission for spins is absent. Then, what could 
initiate the motion of spins? What mechanism would be the origin of pure spin 
superradiance?

There existed a widespread delusion that the thermal Nyquist noise of the 
resonant electric circuit could be the initiating cause of spin rotation. At 
this point, it is necessary to stress that the role of the thermal Nyquist 
noise in producing a fluctuating torque on the magnetization was in detail 
discussed by Bloembergen and Pound in their classic paper{\large$^6$}. They 
showed that the average thermal damping is inversely proportional to the 
sample volume, because of which this damping could be noticeable only for a 
single spin, but for a macroscopic sample of many spins $N\gg 1$ the coil 
thermal damping should be negligibly small. This analysis seems to have been 
completely forgotten by later workers who have proclamed the leading role 
of the coil thermal noise in initiating spin rotation.

To explicitly illustrate the role of the thermal Nyquist noise, one has to 
consider the effective attenuation width (48), including the influence of 
the electromotive force related to the resonance mode of the thermal 
noise{\large$^{19-21}$}. When there are no external transverse magnetic 
fields, that is when $\nu_1=\nu_2=0$, the effective width (48) reads
\begin{equation}
\label{51}
\Gm_{eff} =\Gm_3 + \frac{\nu_0^2}{16\Gm} \left ( 1 -
e^{-\Gm t}\right ) \left ( 1 - e^{-\gm t}\right )^2 \; .
\end{equation}
The electromotive force, corresponding to the resonator thermal noise, can be 
written as
\begin{equation}
\label{52}
\tilde E = \tilde E_0\cos\om t \; .
\end{equation}
The related magnetic field, defined in eqs. (18) and (29), has the amplitude
\begin{equation}
\label{53}
h_0 = \frac{c\tilde E_0}{n A_c\gm} \; .
\end{equation}
From here, it follows that
\begin{equation}
\label{54}
h_0^2 = 8\pi\; \frac{\eta\rho_n\tilde E_0^2}{\gm RN} \; .
\end{equation}
With the definition (37), one gets the quantity
\begin{equation}
\label{55}
\nu_0^2 = \frac{8\Gm_0\tilde E_0^2}{\hbar\gm IRN} \; ,
\end{equation}
which characterizes in eq. (51) the average thermal damping due to the 
Nyquist noise.

First of all, eq. (51) shows that the thermal damping at $t=0$ is exactly zero
because of the temporal factors. The latter become essential only for 
$t>T_2,\tau$. Therefore, the influence of the thermal noise at short times 
should be suppressed by these temporal factors.

Moreover, the quantity (55) is inversely proportional to the number of spins. 
That is, the thermal damping is inversely proportional to $N$, and thus it 
should be negligibly small for a macroscopic sample with $N\gg 1$. This is in 
a complete agreement with the conclusion of Bloembergen and 
Pound{\large$^6$}.

To be more precise, it is possible to explicitly calculate the quantity (55) 
substituting there the square amplitude of the electromotive force due to 
the thermal noise{\large$^7$}, which is \begin{equation}
\label{56}
\tilde E_0^2 = \frac{\hbar\om}{2\pi}\; \gm R \; {\rm coth}\;
\frac{\hbar\om}{2k_BT} \; .
\end{equation}
At radio-frequencies, the inequality
\begin{equation}
\label{57}
\frac{\om}{\om_T} \ll 1 \qquad \left ( \om_T \equiv
\frac{k_BT}{\hbar} \right ) 
\end{equation}
is valid. As a result, eq. (56) simplifies to
\begin{equation}
\label{58}
\tilde E_0^2 \simeq \frac{\hbar}{\pi}\; \gm R\om_T \; .
\end{equation}
Thence, eq. (55) becomes
\begin{equation}
\label{59}
\nu_0^2 = \frac{8\Gm_0\om_T}{\pi IN} \; .
\end{equation}

Expression (51) shows that the thermal damping has to be compared to the
dynamic broadening width $\Gm_3$, which can be of order of $\Gm_2$. When 
the inequality
\begin{equation}
\label{60}
\frac{\nu_0^2}{\Gm_2^2} \ll 1
\end{equation}
holds true, the thermal damping can be safely neglected, playing no role 
in spin relaxation, as compared to the dynamic broadening.

To estimate the factor (59) and, respectively, to check the inequality (60), 
one can accept the values typical of experiments on proton spin 
superradiance{\large$^{11-13}$}. Then, for the filling factor $\eta=0.6$, 
proton density $\rho_n=4\times 10^{22}$ cm$^{-3}$, and the proton magnetogyric 
ratio $\gm_n=2.675\times 10^8$ s$^{-1}$T$^{-1}$, the linewidth (35) is 
$\Gm_0=2.846\times 10^4$ Hz. At temperature $T=0.1$ K, the thermal frequency 
is $\om_T=1.309\times 10^{11}$ Hz. Under the static magnetic field $B_0=1$ T, 
the proton NMR frequency is $\om_0=2.675\times 10^8$ Hz. The ratio 
$\om/\om_T\sim 10^{-3}$ is small. With the transverse relaxation width 
$\Gm_2=1.176\times 10^5$ Hz, one obtains
$$
\frac{\nu_0^2}{\Gm_2^2} = \frac{1.37}{N}\times 10^6 \; .
$$
For a sample volume of about 1 cm$^3$ and $N\sim 10^{23}$, one has 
$\nu_0^2/\Gm_2^2\sim 10^{-17}$. This ratio is so much tiny that, clearly,
the thermal Nyquist noise is not able to play any role in spin relaxation 
in a macroscopic sample. It is only for the number of spins $N\leq 10^6$,
when the thermal noise would be noticeable. When considering macroscopic 
samples, with $N\gg 10^6$, the resonator thermal noise is to be neglected.

\section{Incoherent Stage}

At the initial stage, the main role in starting spin relaxation is played 
by the dynamic broadening due to local spin fluctuations. Omitting in the 
effective width (48) the terms of second order with respect to small 
parameters and neglecting the resonator thermal noise by setting $\bt\ra 0$, 
one has
\begin{equation}
\label{61}
\Gm_{eff} = \Gm_3 + \frac{\nu_2^2}{4\Gm} \left ( 1 - 
e^{-\Gm t}\right ) \; .
\end{equation}
This equation shows that at the very beginning of the process, when $t\ra 0$, 
then $\Gm_{eff}\ra\Gm_3$. That is, the dynamic broadening width $\Gm_3$ 
makes the major contribution to the effective relaxation width, eq. (61).

For asymptotically small times $t\ra 0$, the coupling function
(42) is close to zero, $\al\ra 0$. Then the evolution equations
(49) and (50) can be reduced to the form
\begin{equation}
\label{62}
\frac{dw}{dt} = - 2\Gm_2 w + 2\Gm_3 s^2 \; , \qquad 
\frac{ds}{dt} = - \Gm_1^*( s - \zeta^* ) \; ,
\end{equation}
in which
\begin{equation}
\label{63}
\Gm_1^* \equiv \Gm_1 + \Gm_3 \; , \qquad
\zeta^* \equiv \frac{\Gm_1}{\Gm_1^*}\; \zeta \; .
\end{equation}
The solutions to eqs. (62) are
\begin{equation}
\label{64}
w= w_0 e^{-2\Gm_2 t} + 2\Gm_3 
\int_0^t s^2(t') e^{-2\Gm_2(t-t')} \; dt' \; , \quad
s= \zeta^* + ( s_0 - \zeta^*) e^{-\Gm_1^* t} \; ,
\end{equation}
which demonstrates the character of spin motion at short times $t\ra 0$. At 
this stage, the motion of spins is yet completely incoherent, since, because 
of retardation, the coupling with the resonator has not yet been switched 
on. Assuming that $\Gm_1\ll\Gm_3$, hence $\zeta^*\ll 1$, one may simplify
the solution (64) as
$$
w\simeq w_0 e^{-2\Gm_2 t} + \frac{\Gm_3 s_0^2}{\Gm_2-\Gm_3}
\left ( e^{-2\Gm_3 t} - e^{-2\Gm_2 t}\right ) \; , \qquad
s \simeq s_0 e^{-\Gm_3 t} \; .
$$
The initial {\it incoherent stage} of spin relaxation can also be
called the {\it quantum stage}, as far as the relaxation is caused 
by quantum effects of spin-spin interactions and there are yet no 
collective effects that could lead to the development of coherence.

The duration of the incoherent quantum stage lasts till the {\it
quantum time} $t_q$, when the coupling function (42) reaches the 
value $\Gm_2$, so that
\begin{equation}
\label{65}
\al(t_q) = \Gm_2 \; ,
\end{equation}
and when taking account of collective effects, due to the coupling 
with the resonator, becomes important. It is also easy to notice 
that the difference $\Gm_2-\al s$ in eq. (49) may change its sign
at the quantum time $t_q$, which means that the generation of 
coherent radiation would begin. Combining eqs. (42) and (65) gives 
for the quantum time
\begin{equation}
\label{66}
t_q = \tau \ln\; \frac{g}{g-1} \; ,
\end{equation}
where the notation of the effective coupling parameter
\begin{equation}
\label{67}
g \equiv \frac{\Gm_0\om_0}{\Gm_2\gm} = 2Q\; \frac{\Gm_0}{\Gm_2}
\end{equation}
is introduced. The quantum time (66) is positive only for $g>1$. If $g=1$, 
then $t_q\ra\infty$. This indicates that the quantum stage is finite and may 
change to a collective stage only if the coupling parameter (67) is $g>1$. 
If the spin-resonator coupling is weak, and $g\leq 1$, there exists solely 
the quantum stage, and the collective stage never comes into being. For 
sufficiently strong coupling, eq. (66) gives
\begin{equation}
\label{68}
t_q \simeq \frac{\tau}{g} \qquad (g\gg 1) \; .
\end{equation}

The values of solutions (64) at the end of the quantum stage, i.e.
at $t_q$, can be estimated assuming that $t_q\ll T_2$ and 
$t_q\ll T_3\equiv \Gm_3^{-1}$. Then eqs. (64) yield
\begin{equation}
\label{69}
w(t_q) \simeq w_0 + 2\dlt_3 s^2 \; , \qquad s(t_q)\simeq s_0 \; ,
\end{equation}
where
\begin{equation}
\label{70}
\dlt_3 \equiv \frac{\Gm_3}{g\gm} = 
\frac{\Gm_2\Gm_3}{\Gm_0\om_0} \; .
\end{equation}
This is a small parameter, since $\Gm_0\sim \Gm_3$ while $\Gm_2\ll\om$.
But, no matter how small $\dlt_3$ is, it can be principally important 
if $w_0=0$.

\section{Transient Superradiance}

After the incoherent quantum stage, the coupling function (42) increases,
switching on collective effects resulting in the appearance of spin 
superradiance{\large$^{19-21}$}. An accurate description of this process 
can be made by solving the evolution equations (49) and (50) numerically.
It is also possible to present an explicit analytical description of how 
superradiance develops for the transient stage, when the time is larger than 
the quantum time $t_q$ but yet much smaller than the spin-lattice 
relaxation time $T_1$. Then, the term with $\Gm_1$ in eq. (50) can be 
omitted. If there are no strong external transverse fields, then one should 
set $\nu_1\ra 0$ and $\nu_2\ra 0$. Assuming that the coupling function (42) 
has reached its maximal value, one gets $\al\approx g\Gm_2$. The dynamic
broadening width $\Gm_3$ is not larger than $\Gm_2$. Hence, for sufficiently
large coupling parameter $g$, one may neglect $\Gm_3$ as compared to 
$g\Gm_2\gg\Gm_3$. Under these conditions, eqs. (49) and (50) read
\begin{equation}
\label{71}
\frac{dw}{dt} = -2\Gm_2 ( 1 - gs ) w \; , \qquad
\frac{ds}{dt} = - g \Gm_2 w \; .
\end{equation}
Initial conditions for eqs. (71) should be taken at $t=t_q$.

Equations (71) can be solved exactly{\large$^{19-21}$} yielding
\begin{equation}
\label{72}
w=\left ( \frac{\Gm_p}{g\Gm_2}\right )^2 {\rm sech}^2
\left ( \frac{t-t_0}{\tau_p}\right ) \; , \qquad
s = -\; \frac{\Gm_p}{g\Gm_2}\; {\rm tanh}\left (
\frac{t-t_0}{\tau_p}\right ) + \frac{1}{g} \; .
\end{equation}
Here $\Gm_p$ is a pulse width, $\tau_p=\Gm_p^{-1}$ is a pulse time, 
and $t_0$ is a delay time. These parameters are the integration constants 
that are to be defined from initial conditions $w(t_q)$ and $s(t_q)$ taken
at the boundary of the transient stage, that is at the quantum time $t_q$.
Employing eqs. (69) results in
\begin{equation}
\label{73}
\Gm_p^2 = \Gm_g^2 + \left ( g\Gm_2\right )^2 \left ( w_0 + 2\dlt_3 s_0^2
\right ) \; , \qquad \Gm_g\equiv \Gm_2( 1 - gs_0 ) \; , \qquad
\Gm_p\tau_p = 1 \; ,
\end{equation}
while the delay time is
\begin{equation}
\label{74}
t_0 = t_q +\frac{\tau_p}{2}\; \ln \left |
\frac{\Gm_p - \Gm_g}{\Gm_p+\Gm_g} \right | \; .
\end{equation}
Solutions (72) demonstrate that the coherence intensity $w(t)$ has the 
shape of a burst of width $\tau_p$ and peaked at the delay time $t_0$.

If the spin-resonator coupling $g$ is sufficiently weak or the initial spin 
polarization $s_0$ is sufficiently small, so that $gs_0\leq 1$, than the value 
of the delay time (74) shifts to the quantum region, becoming $t_0\leq t_q$. 
This means that the radiation intensity would not have the shape of a coherent 
burst, but is rather a decreasing function of time, typical of nuclear 
induction. In the case of an external pulse thrusting onto the spins an 
essential initial coherence, such that $g^2w_0 > 1$, the radiation time 
$\tau_p$ is small, $\tau_p <T_2$, which is characteristic of collective 
induction. When $gs_0>1$ and $g^2w_0>1$, the signal of collective induction 
is peaked at a finite delay time $t_0>0$. As is explained in the Introduction, 
collective induction, with coherence being induced by an initial external 
source, is in principal different from superradiance that is a process of 
spontaneously arising coherence.

For solutions (72) to describe a superradiant burst, the requirements 
discussed in the Introduction must be satisfied. In order that the 
initial coherence, given by $w_0$, would not essentially influence 
the pulse width $\Gm_p$, it should be that
\begin{equation}
\label{75}
g^2w_0 < 1 \; .
\end{equation}
This guaranties a basically spontaneous character of the process. Then, 
the pulse time $\tau_p$ is to be sufficiently short and the delay time 
finite, 
\begin{equation}
\label{76}
\tau_p < T_2 \; , \qquad t_q < t_0 < \infty \; .
\end{equation}
The second of these inequalities yields
\begin{equation}
\label{77}
gs_0 > 1 \; , \qquad w_0 + 2\dlt_3 s_0^2 > 0 \; .
\end{equation}
Combining all conditions (75) to (77), one sees that there exist two types of 
transient spin superradiance, whose classification, taking account of the 
smallness of the parameter $\dlt_3\ll 1$, can be accomplished as follows:

\vskip 2mm

(i) {\it Triggered superradiance}, when
$$
gs_0 > 1 +\sqrt{1 -g^2 w_0} \; , \qquad w_0 \neq 0 \; .
$$
Here, a weak external pulse imposes an initial coherence on spins, but, 
being rather weak, this pulse plays just the role of a trigger, while 
the following development of coherence is mainly governed by internal 
properties.

\vskip 2mm

(ii) {\it Pure superradiance}, when
$$
gs_0 > 2 \; , \qquad w_0 = 0 \; .
$$
This is a purely self-organized process, with no coherence prescribed
from external sources.

In the case of pure spin superradiance, the characteristic quantities
(73) and (74), having regard to $\dlt_3\ll 1$, become 
$$
\Gm_p \simeq (g s_0 - 1) \Gm_2 \left [ 1 + \left (
\frac{gs_0}{gs_0 -1}\right )^2 \dlt_3 \right ] \; , \qquad
\tau_p \simeq \frac{T_2}{gs_0-1} \left [ 1 - \left (
\frac{gs_0}{gs_0-1}\right )^2 \dlt_3 \right ] \; ,
$$
\begin{equation}
\label{78}
t_0 \simeq \frac{\tau}{g} + \frac{T_2}{2(gs_0-1)} \; 
\ln\left | \frac{2}{\dlt_3} \left ( 1 - \; \frac{1}{g s_0}
\right )^2 \right | \; .
\end{equation}
Although the parameter $\dlt_3$ is small, it cannot be neglected, since its 
value is crucial for defining the delay time $t_0$. As is seen, if 
$\dlt_3\ra 0$, then $t_0\ra\infty$. The parameter $\dlt_3$, according to the 
notation (70), is proportional to the dynamic broadening width $\Gm_3$ due 
to local spin fluctuations. As far as the latter fluctuations are the main 
cause of starting spin relaxation, it is not surprising that the related 
broadening defines the delay time.

From eqs. (72) it follows that the maximum of the coherent burst occurs
at $t=t_0$, when
\begin{equation}
\label{79}
w(t_0) = \left ( s_0 -\; \frac{1}{g} \right )^2 \; , \qquad 
s(t_0)= \frac{1}{g} \; .
\end{equation}
For longer times $t\gg t_0$, the superradiant signal exponentially 
diminishes, and the spin polarization returns to an almost inverted value,
\begin{equation}
\label{80}
w \simeq 4 w(t_0) e^{-2\Gm_p t} \; , \qquad 
s \simeq - s_0 + \frac{2}{g} \; .
\end{equation}
The larger is the spin-resonator coupling $g$, the more complete is the 
spin inversion. The effect of superradiant spin inversion can be employed
for the ultrafast repolarization of nuclear targets used in scattering 
experiments{\large$^{14,29}$}.

The shape of a superradiant pulse, described by $w(t)$ in eqs. (72), 
is in very good agreement with experiments{\large$^{11-13}$}. Thus, 
the measured signal, presented in Fig. 2, ideally fits the calculated 
function $w(t)$, as is discussed in the papers{\large$^{11,12}$}. 
A concrete relation between the measured intensity of the signal and 
$w(t)$ will be explained in Sec. 10.

It is worth emphasizing that the radiation intensity and radiation
time of the studied pulse have the properties typical of a superradiant 
burst. To illustrate this, it is sufficient to recall that the initial 
inverted spin polarization $s_0=N_+/N$ is the ratio of the number of
inverted spins to their total number in the sample. Inverted spins play 
the role of radiators, whose number is $N_+$. For large spin-resonator 
coupling $g\gg 1$, one has $w(t_0)\sim s_0^2$ and $\tau_p\sim 1/s_0$,
which is clear from the consideration above. Therefore
$$
w(t_0) \sim N_+^2 \; , \qquad \tau_p \sim N_+^{-1} \; ,
$$
which is characteristic of superradiance.

When there is no constant pumping by means of dynamic nuclear 
polarization, there appears one dominant superradiant burst given by 
eqs. (72). In an inhomogeneous sample, after the first burst, secondary 
maser oscillations can arise, whose accurate description requires 
numerical investigation{\large$^{15-17,30-32}$}.

It is interesting to mention that the signals of nuclear spin echo can
also be essentially amplified by the presence of a resonant electric 
circuit coupled with a spin system, which has been observed in 
experiment{\large$^{33}$}.

\section{Pulsing Superradiance}

Another regime of spin superradiance can be achieved if the
inversion of nuclear spins is constantly supported by a pumping 
mechanism, for instance, by dynamic nuclear polarization. Then, 
a series of superradiant bursts can be produced, as is 
experimentally observed{\large$^{8-10}$}. Such a regime, with a series of
superradiant pulses that may repeatedly arise during a rather long 
time, can be termed pulsing superradiance. The theoretical description
of this regime is based on the evolution equations (49) and (50), which 
require numerical computations{\large$^{17,22,24}$}. However, for the 
late stage, when $t\gg T_2,\tau$, and the system is close to a stationary
state, it is possible to solve the problem analytically.

At the late stage, when $t\gg \tau$, the coupling function (42) becomes 
$\al=g\Gm_2$. As is explained in Section 5, the resonator thermal noise can 
always be neglected, that is $\bt\ra 0$. When there are no strong external 
transverse fields, then $\nu_1\ra 0$ and $\nu_2\ra 0$. For the simplicity, 
the dynamic broadening width $\Gm_3$ can also be neglected as compared to a 
large $g\Gm_2$. Then the evolution equations (49) and (50) can be written as 
a two-dimensional dynamical system
\begin{equation}
\label{81}
\frac{dw}{dt} = v_1 \; , \qquad \frac{ds}{dt} = v_2 \; ,
\end{equation}
with the velocity fields
\begin{equation}
\label{82}
v_1 = -2\Gm_2( 1 - gs ) w \; , \qquad
v_2 = - g\Gm_2 w - \Gm_1^* ( s - \zeta ) \; .
\end{equation}
In the presence of dynamic nuclear polarization, $\Gm_1^*$ is a pumping 
rate and $\zeta>0$ is a pumping parameter.

An accurate theoretical analysis of eqs. (81) is given below. The velocity 
fields are taken in the simplified form (82). This is done for pedagogical
reasons, in order not to become entangled into cumbersome formulas, but for
emphasizing the principal ideas and techniques, on which such an analysis is 
based. Using the same methods, one could accomplish an analysis of the
evolution equations (49) and (50), including the dynamic broadening and
external transverse fields, which would result in much more intricate 
equations.

An important quantity characterizing local stability of motion is the local
expansion rate{\large$^{34}$}\begin{equation}
\label{83}
\Lambda(t) \equiv \frac{1}{t}\; {\rm Re} 
\int_0^t {\rm Tr}\hat X(t')\; dt'  \; ,
\end{equation}
whose definition involves the Jacobian expansion matrix $\hat X$ that in the
considered case is
\begin{eqnarray}
\hat X(t) \equiv \left [ \begin{array}{cc}

\frac{\prt v_1}{\prt w} & \frac{\prt v_1}{\prt s} \\
\nonumber

\frac{\prt v_2}{\prt w} & \frac{\prt v_2}{\prt s}
\end{array} \right ] \; .
\end{eqnarray}
If, at the moment $t$, the rate (83) is positive, $\Lambda(t)>0$, this means
that the phase volume of the dynamical system expands, while if $\Lambda(t)<0$,
then the phase volume contracts. The eigenvalues of the expansion matrix 
$\hat X$ are the local characteristic exponents
$$
X^\pm = -\; \frac{1}{2} \left \{ \Gm_1^* + 2\Gm_2 ( 1 - gs ) \pm
\sqrt{ \left [ \Gm_1^* - 2\Gm_2 ( 1 - gs )\right ]^2 - 8g^2 \Gm_2^2 w} 
\right \} \; ,
$$
whose real parts define the local Lyapunov exponents
$$
\lbd^\pm(t) \equiv {\rm Re}\; X^\pm(t) \; .
$$
The signs of the latter show whether the motion at the moment $t$ is stable
or not.

In the theory of dynamical systems{\large$^{35}$}, one usually considers 
asymptotic stability related to the limit $t\ra\infty$. The stationary, or 
fixed, points are given by the zeros of the velocity fields. The equations 
$v_1=v_2=0$, with the velocities (82), yield two stationary solutions: one 
fixed point is
\begin{equation}
\label{84}
w_1^* =0 \; , \qquad s_1^* = \zeta 
\end{equation}
and another is
\begin{equation}
\label{85}
w_2^* =\frac{\Gm_1^*}{g^2\Gm_2}\left ( g\zeta - 1 \right ) \; , \qquad
s_2^* = \frac{1}{g} \; .
\end{equation}
As is clear, among several solutions, only those could have sense whose 
values are in the region of validity of the considered functions. This
region, as follows from eqs. (22) and (23), is limited by the inequalities
$$
0 \leq w \leq 1 \; , \qquad -1 \leq s \leq 1 \; .
$$
Thence, it is evident that the fixed point (85) may have sense only if 
$g\zeta\geq 1$. When $g\zeta=1$, both fixed points (84) and (85) coincide.
On the manifold of system parameters, the value $g\zeta=1$ is termed a
bifurcation point.

Because of the existence of two stationary solutions, the limit
$$
\Lambda^* \equiv \lim_{t\ra\infty}\Lambda(t)  = {\rm Re}\;
\lim_{t\ra\infty}{\rm Tr}\; \hat X(t)
$$
of the local expansion rate (83), for which $\Lambda^*=\lbd^+ +\lbd^-$,
with the Lyapunov exponents
$$
\lbd^\pm \equiv {\rm Re}\; \lim_{t\ra\infty} X^\pm(t) \; ,
$$
possesses two different values. The local expansion rates at the
first and second fixed points, respectively, are
\begin{equation}
\label{86}
\Lambda^*_1 = -\Gm_1^* - 2\Gm_2 ( 1 - g\zeta )\; , \qquad
\Lambda_2^* = -\Gm_1^* \; .
\end{equation}
According to the principle of minimal expansion{\large$^{34}$}, the smaller 
the local expansion rate, the more stable is a dynamic state. The values (86) 
show that when $g\zeta<1$, the fixed point (84) is more stable, while when 
$g\zeta>1$, the stationary solution (85) becomes more stable.

The limits $X^\pm\equiv\lim_{t\ra\infty}X^\pm(t)$ of the characteristic 
exponents, at the corresponding fixed points, are
$$
X^+_1 = -\Gm_1^* \; , \qquad X_1^- = - 2\Gm_2 (1 - g\zeta) \; ,
$$
\begin{equation}
\label{87}
X_2^\pm = -\; \frac{\Gm_1^*}{2}\left [ 1 \pm \sqrt{ 1 - 8\; 
\frac{\Gm_2}{\Gm_1^*} ( g\zeta -1 )} \right ] \; .
\end{equation}
The related real parts $\lbd^\pm={\rm Re}X^\pm$ define the Lyapunov 
exponents. Analysing eqs. (87) yields the following classification 
of the fixed points (84) and (85).

When $g\zeta<1$, then the Lyapunov exponents for the stationary solution
(84) are negative, $\lbd_1^\pm<0$. This tells that the fixed point (84) 
is a stable node. For the stationary solution (85), one has $\lbd_2^+<0$ 
but $\lbd_2^->0$, which means that the fixed point (85) is a saddle. 
Hence, for $g\zeta<1$, spin dynamics tends to the stationary solution 
(84). The relaxation rates are given by $\lbd_1^\pm=X_1^\pm$ defined in 
eqs. (87), from which it follows that the spin motion is incoherent.

In the case $g\zeta=1$, the stationary solution is unique, since the 
fixed points (84) and (85) coincide. The related Lyapunov exponents are 
also the same: $\lbd_1^+=\lbd_2^+<0$, $\lbd_1^-=\lbd_2^-=0$. The value
$g\zeta=1$ is associated with a bifurcation point.

With the increasing pumping, when
$$
1 < g\zeta < 1 +\frac{\Gm_1^*}{8\Gm_2} \; ,
$$
one gets $\lbd_1^+<0$, $\lbd_1^->0$, while $\lbd_2^\pm<0$. This indicates
that the features of the fixed points have been changed $-$ the fixed point 
(84) is now a saddle, while that of eq. (85) has turned to a stable node. 
For realistic nuclear spins, $\Gm_1^*\ll\Gm_2$. Therefore, the considered
region of $g\zeta$ is very narrow. The relaxation rates are close to 
$\lbd_2^\pm\approx -\frac{1}{2}\Gm_1^*$; hence the spin motion is to be 
incoherent. The coherence intensity $w_2^*$, given in eq. (85), although is 
not zero exactly, but, since $\Gm_1^*\ll\Gm_2$ and $g>1$, is very small, 
not much differing from zero.

Increasing the pumping further, so that
$$
g\zeta > 1 + \frac{\Gm_1^*}{8\Gm_2} \; ,
$$
changes the picture qualitatively. Then the fixed point (84) continues to 
be a saddle, since $\lbd_1^+<0$, $\lbd_1^->0$, but the fixed point (85) 
turns into a stable focus, as far as its characteristic exponents become 
complex,
\begin{equation}
\label{88}
X_2^\pm = -\; \frac{\Gm_1^*}{2} \pm i\om_\infty \; ,
\end{equation}
with the asymptotic frequency
\begin{equation}
\label{89}
\om_\infty \equiv \frac{\Gm_1^*}{2}\; \sqrt{8\; \frac{\Gm_2}{\Gm_1^*}\;
(g\zeta -1 ) -1} \; .
\end{equation}
This is the regime of pulsing spin superradiance, when there appears a 
long series of superradiant bursts lasting about the time $T_1^* = 
(\Gm_1^*)^{-1}$. At the intermediate stage, this pulsing is not periodic, 
but becomes approximately periodic as time increases. The asymptotic 
frequency (89) defines the time interval $T_\infty\equiv 2\pi/\om_\infty$
between superradiant pulses at the late stage, when the spin evolution is
close to the stationary solution (85). This interpulse time for sufficiently
strong pumping, with $g\zeta\gg 1$, is
\begin{equation}
\label{90}
T_\infty \equiv \frac{2\pi}{\om_\infty} \simeq \pi  \;
\sqrt{\frac{2T_1^* T_2}{g\zeta} } \; .
\end{equation}
The time (90), by varying the related parameters, can be varied widely. 
For example, keeping in mind the parameters typical of the proton 
spins{\large$^{11-13}$}, one has $\Gm_0=2.85\times 10^4$ Hz, $\om_0=2.675
\times 10^8$ Hz, $\Gm_2=1.18\times 10^5$ Hz. With the quality factor $Q=100$, 
the ringing width is $\gm=1.34\times 10^6$ Hz. The spin-resonator coupling 
(67) is $g=48$. For the pumping characteristics $\zeta=1$ and $\Gm_1^*=10$ Hz, 
the interpulse time (90) is $T_\infty=0.59\times 10^{-3}$ s. The number of 
pulses can be estimated as $T_1^*/T_\infty\sim 100$. A detailed description 
of the regime of pulsing spin superradiance, for arbitrary times $t>0$, can 
be accomplished with the help of numerical calculations{\large$^{17,22,24}$}. 
The phenomenon of pulsing spin superradiance may be employed for creating 
pulsing spin masers{\large$^{22}$}.

A regime, similar to pulsing spin superradiance, can also be realized without 
dynamic nuclear polarization. For this purpose, one could change in the 
following way the experimental setup used for realizing the transient spin 
superradiance. In the latter regime, as is described in Section 7, if the 
spin-resonator coupling $g$ is sufficiently strong, a sharp superradiant 
burst occurs at the delay time $t_0$. After the time $t_0+\tau_p$, the spin 
polarization, according to eq. (80), becomes practically inverted, provided 
$g\gg 1$. If at this moment, one inverts the static magnetic field $B_0$ or 
acts on spins by a resonant $\pi$-pulse, or just turns the sample $180^0$ 
about an axis perpendicular to $B_0$, then again a strongly nonequilibrium 
state of almost completely inverted spins is prepared.  After the time $t_0$, 
counted from the moment when the newly inverted state is created, another 
superradiant burst will arise. Then, one could again either invert the 
static magnetic field or invert the magnetization by a resonant 
$\pi$-pulse, or turn the sample to arrange a novel nonequilibrium 
inverted state of spins. After this, one more superradiant burst will appear. 
Such a procedure can be repeated as many times as necessary for producing a 
required number of sharp superradiant pulses. The achieved regime may be named 
{\it punctuated spin superradiance}. The principal difference of this regime 
from the pulsing spin superradiance is the possibility of controlling the number
of pulses as well as the interpulse time. The term {\it punctuation} here 
implies the possibility of varying the time intervals between pulses and to form
groups of pulses, containing different numbers of the latter. In that way, it 
is feasible to compose a code, similar to the Morse alphabet. Hence, punctuated 
spin superradiance can be used for processing information, which may be 
employed, for instance, in quantum computers.

\section{Hyperfine Interactions}

Real solids, in addition to nuclei, always contain electrons. It is therefore
important to understand the influence of hyperfine spin-spin interactions 
between nuclei and electrons on nuclear spin superradiance.

A system of nuclear and electronic spins is described by the Hamiltonian
\begin{equation}
\label{91}
\hat H = \hat H_{nuc} + \hat H_{ele} + \hat H_{hyp} \; ,
\end{equation}
in which the nuclear term $\hat H_{nuc}$ is given by eq. (1). The electronic
spin Hamiltonian is
\begin{equation}
\label{92}
\hat H_{ele} = -\; \frac{1}{2} \sum_{i\neq j} J_{ij} \hat{\bf S}_i \cdot
\hat{\bf S}_j + \hbar \gm_e \sum_i {\bf B}\cdot\hat{\bf S}_i \; ,
\end{equation}
where $J_{ij}$ is an exchange interaction potential, $\hat{\bf S}_i$ is an
electron spin, and $\gm_e$ is the electron magnetogyric ratio. The part 
of the Hamiltonian  (91) describing hyperfine interactions is
\begin{equation}
\label{93}
\hat H_{hyp} = A \sum_i \hat{\bf S}_i \cdot \hat{\bf I}_i + \frac{1}{2}
\sum_{i\neq j} \sum_{\al\bt} A_{ij}^{\al\bt}\hat S_i^\al\hat I_j^\bt \; .
\end{equation}
Here the first term is the Fermi contact hyperfine interaction of nuclei 
with $s$-electrons, characterized by the energy
\begin{equation}
\label{94}
A = \frac{8\pi}{3}\; \hbar^2 \gm_e\gm_n |\psi(0)|^2 \; ,
\end{equation}
with $\psi({\bf r})$ being the electron wave function. The second 
term in eq. (93) presents the dipolar hyperfine interactions, with
\begin{equation}
\label{95}
A_{ij}^{\al\bt} = -\hbar^2 \; \frac{\gm_e\gm_n}{r_{ij}^3}
\left ( \dlt_{\al\bt} - 3 n_{ij}^\al n_{ij}^\bt \right ) \; .
\end{equation}

The consideration of the problem with the Hamiltonian (91) can be accomplished 
by employing the same methods as have been detailed in the previous sections. 
The main technical difference is that the consideration becomes more cumbersome,
as now it is necessary to deal with six evolution equations for spins, three 
of which are for nuclear spins and three other, for electronic 
spins{\large$^{23,24}$}. The seventh equation is the Kirchhoff equation (19), 
in which the magnetization density now is a sum
\begin{equation}
\label{96}
M_x = \frac{\hbar}{V} \sum_i \left ( \gm_n\; <\hat I_i^x>\; - \; \gm_e\;
<\hat S_i^x>\right )
\end{equation}
of the terms due to nuclear and electronic spins. The evolution 
equations can be again solved by invoking the scale separation 
approach{\large$^{19-21,24,27}$}. The most important conclusions resulting 
from the presence of hyperfine interactions are as follows.

Local spin fluctuations of electrons lead to the increase of the nuclear
dynamic broadening, which results in the sum
\begin{equation}
\label{97}
\tilde\Gm_3 = \Gm_3 + \Gm_3'
\end{equation}
of the widths caused by nuclear dipolar interactions, $\Gm_3$, and by 
electron-nuclear hyperfine interactions, $\Gm_3'$. The relation between
these widths is given by the ratio
\begin{equation}
\label{98}
\frac{\Gm_3'}{\Gm_3} = \frac{\rho_e\mu_e}{\rho_n\mu_n} \; ,
\end{equation}
in which $\rho_e$ and $\rho_n$ are the electron and nuclear densities, while 
$\mu_e\equiv\hbar\gm_e S$ and $\mu_n\equiv\hbar\gm_n I$ are the electron and 
nuclear magnetic moments, respectively. When the electron and nuclear densities 
are close to each other, then, because $\mu_e/\mu_n\sim 10^3$, the ratio (98) 
can be large compared to unity. Therefore, the width (97) can be essentially 
increased by the hyperfine interactions. This results in the shortening of the 
delay time $t_0$ of a superradiant burst, as well as in shortening the pulse 
time $\tau_p$.

In the electronic subsystem, there may appear long-range magnetic order 
due to the exchange interaction of electronic spins. If so, the NMR 
frequency shifts to the value
\begin{equation}
\label{99}
\om_n = \om_0 + \frac{A}{\hbar}\; S m_z \; ,
\end{equation}
where $m_z$ is the $z$-projection of the average electronic magnetization 
normalized to unity, $A$ being the parameter (94). Then, the frequency (99)
will enter all formulas instead of $\om_0$.

Long-range magnetic order of electrons leads to the renormalization of the
coupling with the resonator according to the rule
\begin{equation}
\label{100}
\tilde g = g \left ( 1 + \frac{\rho_e\mu_e A}{\rho_n\mu_n\hbar\om_n}\;
m_z\right ) \; .
\end{equation}
This may result because of the large ratio $\mu_e/\mu_n\sim 10^3$ in 
an essential increase of the coupling (100), as compared to $g$. The
electronic subsystem plays the role of an additional resonator, which
enhances the effective coupling of nuclear spins with the resonant
electric circuit{\large$^{24}$}.

\section{Radiation Intensity}

To complete the treatment, it is worth emphasizing once again why, 
actually, the studied phenomenon can be called spin superradiance. An
ensemble of coherently moving nuclear spins, as is clear, generates the
magnetodipole radiation with the total intensity
\begin{equation}
\label{101}
I(t) = \frac{2}{3c^3} \left |\ddot{\bf M}(t)\right |^2 \; ,
\end{equation}
where
$$
{\bf M}(t) = \hbar \gm_n \sum_i < {\bf I}_i(t)>
$$
is the total magnetization of nuclei. The quantity of interest is the 
radiation intensity averaged over fast oscillations. For the intensity
(101), this yields
\begin{equation}
\label{102}
\overline I(t) = \frac{2}{3c^3} \; \mu_n^2 \om_0^4 N^2 w(t) \; ,
\end{equation}
where, as always, $\mu_n\equiv\hbar\gm_n I$ is the nuclear magnetic moment.
As is seen, the radiation intensity is proportional to the number of spins 
squared, which is characteristic of superradiance.

Though coherently moving spins do produce superradiance, the radiation 
intensity (102) is not high. At the peak of a superradiant burst, $w(t_0)$ 
is given by eq. (79). For $g\gg 1$ and $s_0\approx 1$, one has 
$w(t_0)\approx 1$. Then the maximal radiation intensity, for proton spins, 
with $\mu_n=1.411\times 10^{-26}JT^{-1}$, $\om_0=2.675\times 10^8$ Hz, and 
$N\approx 10^{23}$, is $\overline I(t_0)\approx 2.5\times 10^{-5}$ W.

Despite so weak a radiation intensity, it can be easily detected. This is 
because what is directly measured is the power of current
\begin{equation}
\label{103}
P(t) = Rj^2(t) \; ,
\end{equation}
which is generated in a coil by the radiating spins. Using the relation of 
the induced current with the resonator magnetic field (14), one gets
$$
j^2(t) = \frac{V_c}{4\pi L}\; H^2(t) \; .
$$
The field $H$ can be found from eq. (41). Setting in this equation $\bt=0$, 
i.e. neglecting thermal noise, and averaging over fast oscillations gives
$$
\overline{H^2(t)} = \frac{2}{\gm_n^2}\; \al^2(t) w(t) \; .
$$
Then, for the averaged current power (103), one finds
\begin{equation}
\label{104}
\overline P(t) = g\Gm_2 I\hbar\om N \left ( 1 - e^{-\gm t}\right )^2
w(t) \; .
\end{equation}
This expression contains, as compared to eq. (102), an additional dependence 
on the feedback retardation. But for $t\gg\tau$, the radiation intensity 
(102) and the current power (104) differ one from another only by a numerical 
factor
\begin{equation}
\label{105}
\frac{\overline P(t)}{\overline I(t)} \simeq 
\frac{3g\Gm_2\lbd^3}{16\pi^2\Gm_0 V_c} = \frac{3Q\lbd^3}{8\pi^2 V_c} \; ,
\end{equation}
in which $\lbd=2\pi c/\om$ is the radiation wavelength. For an NMR frequency 
$\om=2.675\times 10^8$ Hz, one has $\lbd=0.71\times 10^3$ cm. The factor 
(105) can reach rather large values. Thus, if $Q=100$ and $V_c=10$ cm$^3$, 
this factor is of order $10^8$. This is why even a low radiation intensity 
can be easily measured.

\section{Applications}

Nuclear spin superradiance is collective spontaneous radiation by nuclear 
spins at a frequency of nuclear magnetic resonance. This phenomenon is an 
analog of atomic superradiance occurring at optical frequencies. Being 
a novel coherent phenomenon at NMR frequencies, it may find various 
applications, among which it is possible to suggest the following:

\vskip 2mm

(1) {\it Investigation of materials characteristics}

\vskip 1mm

Spin relaxation in materials is commonly employed for studying relaxation 
parameters describing intrinsic properties of matter. Nuclear spin 
superradiance is accomplished by a self-organized coherent spin relaxation, 
which is drastically different from other types of spin relaxation. Another 
kind of relaxation may provide additional information on the properties of 
materials. Thus, the main mechanism originating pure spin superradiance is 
the existence of local spin fluctuations caused by the interaction of nuclear 
spins with each other, by means of dipolar forces, and with electronic spins, 
through hyperfine forces. Therefore, studying specific features of pure spin 
superradiance supplies information about these local fluctuations.

\vskip 2mm

(2) {\it Fast repolarization of targets}

\vskip 1mm

The relaxation of nuclear spins, in the process of superradiance, happens 
much faster than the usual spin-dephasing time $T_2$. If the initial spin 
polarization is sufficiently high and the spin-resonator coupling is 
sufficiently strong, spins, after a superradiant burst, change their 
orientation becoming almost completely inverted. Such an ultrafast inversion 
of spins can be employed for quick repolarization of polarized solid-state 
targets used is scattering experiments.

\vskip 2mm

(3) {\it Construction of spin masers}

\vskip 1mm

A superradiant spin system is a source of coherent radiation at 
radiofrequencies. Being a source of coherent radiation, it is analogous to 
lasers operating at optical frequencies. Spin masers could find applications 
similar to those of optical lasers. A spin maser can function in a transient 
regime, emitting a single burst typical of pure or triggered spin superradiance, 
and also it can work in a lasting regime based on pulsing spin superradiance.

\vskip 2mm

(4) {\it Creation of sensitive detectors}

\vskip 1mm

Spin superradiance can be triggered by a very weak external pulse of intensity 
corresponding to $w_0$. The latter depends exponentially on the delay time 
$t_0$. Therefore, measuring the delay time of a superradiant burst provides 
an accurate evaluation of the triggering intensity. Then, triggered spin 
superradiance may serve as a sensitive mechanism for detecting weak external 
signals.

\vskip 2mm

(5) {\it Methods of information processing}

\vskip 1mm

In the regime of punctuated superradiance, the intervals between superradiant 
bursts as well as the number of pulses can be regulated. This makes it feasible 
to compose a kind of the Morse Code alphabet and, respectively, to develop a 
technique of processing information. Such a method of information processing
could be used in quantum computers.

\vskip 2mm

In conclusion, it is necessary to note that the developed theory, as well 
as possible applications, are appropriate not only for nuclear spins but, in 
general, for any spin system. For instance, all this can immediately be 
extended to electronic spins. The related phenomenon is {\it electron spin 
superradiance}. Other types of materials could be the so-called molecular 
magnets formed of magnetic molecules. The latter often have high spins, thus 
possessing several quantum transitions. In that case, the resonant electric 
circuit has to be tuned to one of the admissible transition frequencies. 
The resulting effect would be {\it molecular spin superradiance}, whose 
intensity could be much higher than that of nuclear spins. Nuclear spin 
superradiance is just a representative from a class of phenomena called 
{\it spin superradiance}. These phenomena, being realized with different 
materials, should exhibit a rich variety of specific features that could 
be not solely interesting but also useful for various applications.

\newpage

\begin{center}
{\bf Figure Captions}
\end{center}

{\bf Fig. 1}. Typical experimental setup for detecting spin superradiance.
The region surrounded by the dashed curve signifies a cryostat. Among the
two coils below, the left one plays the role of antenna, and the right one 
is a part of the resonant electric circuit. The studied sample is shown as 
a dark bar inside the resonant coil. The upper left block is an oscilloscope, 
and the lower one is a plotter.

\vskip 5mm

{\bf Fig. 2}. Voltage signal corresponding to a superradiant pulse as a 
function of time, measured in units of $10^{-7}$ s, for two initial spin 
polarizations, $s_0=0.52$ (lower curve) and $s_0=0.57$ (upper curve).

\vskip 5mm

{\bf Fig. 3}. Orientation of the coordinate axes with respect to the 
sample that is inserted into the coil of a resonant electric circuit.

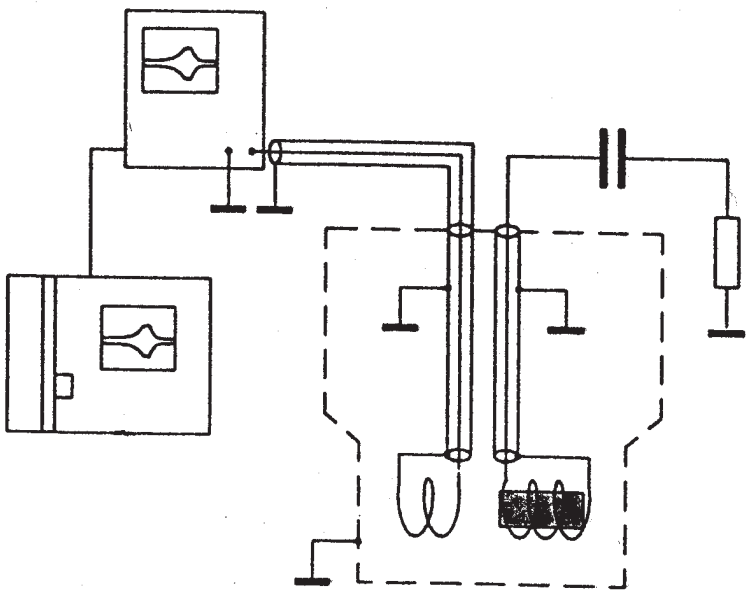
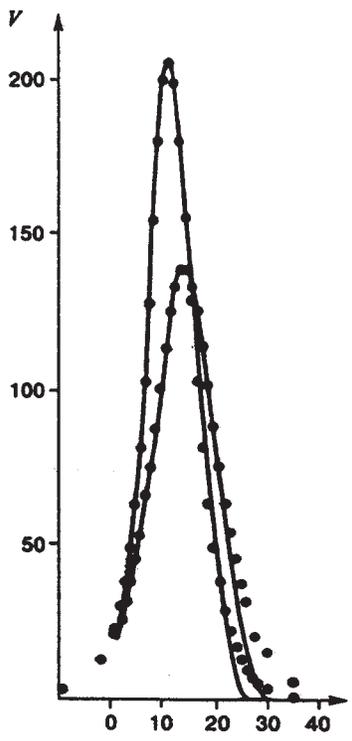
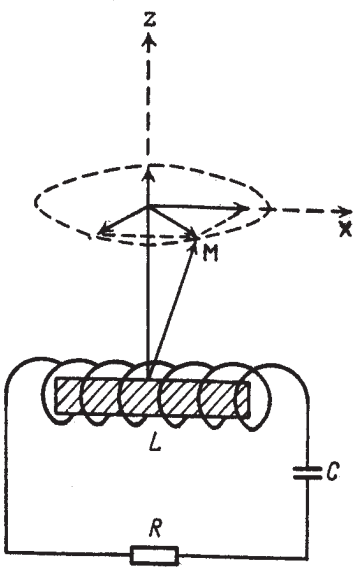
\end{document}